\documentclass[aps, prc,nofootinbib,twocolumn, superscriptaddress]{revtex4-2}

\usepackage[utf8]{inputenc}
\usepackage{graphicx}

\usepackage{algorithm}
\usepackage{algpseudocode}
\usepackage{amsthm, amsfonts, amsmath, mathrsfs}
\usepackage{bbm}
\usepackage{slashed}
\usepackage{physics}
\usepackage{xspace}
\usepackage{enumitem}
\setlist[enumerate]{itemsep=0.01cm}

\usepackage{orcidlink}

\usepackage{comment}
\usepackage{glossaries}
\setacronymstyle{long-short}
\glsdisablehyper

\usepackage{hyperref}
\hypersetup{colorlinks=true,
    linkcolor=magenta,
    citecolor=blue,
    urlcolor=cyan,
    pdfpagemode=FullScreen,}

\newacronym{pionless}{EFT$_\slashed \pi$}{pionless effective field theory}
\newacronym{e1}{E1}{electric dipole}
\DeclareMathOperator*{\SumInt}{%
\mathchoice%
  {\ooalign{$\displaystyle\sum_f$\cr\hidewidth$\displaystyle\int$\hidewidth\cr}}
  {\ooalign{\raisebox{.14\height}{\scalebox{.7}{$\textstyle\sum$}}\cr\hidewidth$\textstyle\int$\hidewidth\cr}}
  {\ooalign{\raisebox{.2\height}{\scalebox{.6}{$\scriptstyle\sum$}}\cr$\scriptstyle\int$\cr}}
  {\ooalign{\raisebox{.2\height}{\scalebox{.6}{$\scriptstyle\sum$}}\cr$\scriptstyle\int$\cr}}
}

\usepackage{ulem}
 \newcommand\redsout{\bgroup\markoverwith{\textcolor{red}{\rule[0.5ex]{2pt}{1.4pt}}}\ULon}

 \renewcommand{\emph}[1]{\textit{#1}}

\begin{document}

\title{Optimized binning for response function reconstruction via Chebyshev expansions}

\author{Immo C. Reis\,\orcidlink{0009-0008-3566-6095}}
 \email{ireis@uni-mainz.de}
\affiliation{Institut f\"ur Kernphysik and PRISMA$^+$ Cluster of Excellence, Johannes Gutenberg-Universit\"at, 55128 Mainz, Germany}
\author{Joanna E. Sobczyk\,\orcidlink{0000-0003-4698-9339}}
\email{joanna.sobczyk@chalmers.se}
\affiliation{Institut f\"ur Kernphysik and PRISMA$^+$ Cluster of Excellence, Johannes Gutenberg-Universit\"at, 55128 Mainz, Germany}
\affiliation{Department of Physics, Chalmers University of Technology, SE-412 96 G\"oteborg, Sweden}
\author{Sonia Bacca\,\orcidlink{0000-0002-9189-9458}}
\email{sbacca@uni-mainz.de}
\affiliation{Institut f\"ur Kernphysik and PRISMA$^+$ Cluster of Excellence, Johannes Gutenberg-Universit\"at, 55128 Mainz, Germany}
\affiliation{Helmholtz-Institut Mainz, Johannes Gutenberg Universit\"at Mainz, D-55099 Mainz, Germany}

\begin{abstract}
We propose an optimized histogram binning strategy to reconstruct nuclear response functions via the Chebyshev expansion bound-state method.
Our approach employs a stochastic regularization of the density of states to define adaptive, equal-area bins. Using the deuteron solved in a harmonic-oscillator basis with a chiral interaction, we benchmark on dipole and longitudinal responses, obtaining excellent agreement with exact theory and experiment. This general framework readily extends to other many-body systems and opens the door to new \textit{ab initio} calculations of lepton–nucleus cross sections in medium-mass nuclei.

\end{abstract}

\maketitle

\newpage

\section{Introduction}
\label{sec:introduction}
The interaction of nuclei with electroweak probes has long been pivotal for elucidating nuclear structure and fundamental forces. Interest in these processes has surged with the advent of long-baseline neutrino experiments—such as DUNE and Hyper-Kamiokande~\cite{DUNE:2015lol,10.1093/ptep/ptv061}—that infer oscillation parameters from neutrino–nucleus scattering. Robust interpretation of their data hinges on precise modeling of lepton–nucleus cross-sections, which depend on nuclear response functions encoding the electroweak dynamics of the target nucleus.

From a theoretical standpoint, computing electroweak response functions within \textit{ab initio} methods using interactions derived in an effective field theory~\cite{RevModPhys.81.1773,MACHLEIDT20111,RevModPhys.92.025004} remains a formidable challenge. Most \textit{ab initio} methods are tailored to bound-state problems, employing truncated basis expansions in Hilbert spaces optimized for discrete spectra. Modeling inclusive electroweak reactions is complicated by the need to incorporate nuclear continuum states.
Integral transform methods~\cite{efros2007,Carlson_1992,Lovato_2016} recast the continuum problem as a bound-state calculation, but recovering the original response requires inverting the transform—an ill-posed procedure for which rigorously quantifying theoretical uncertainties remains challenging.

Recently, a histogram‐based method to compute integral transforms via Chebyshev expansions was introduced~\cite{sobczykroggero}. This approach yields the response in narrow energy bins, and allows for a rigorous estimate of upper and lower bounds in each bin. This promises a transparent uncertainty quantification without ill‐posed inversion. The challenge though is to find a systematic approach to assign the bin width in the energy range of interest.

In this work, we devise a binning strategy based on a stochastic estimation of the regularized density of states.
We implement this idea on a simple system, namely the deuteron, where we can benchmark using the exact diagonalization, and compute electromagnetic response functions for which precise theoretical and experimental electromagnetic data exist.
We demonstrate that our implementation reproduces known results to very good accuracy across different kinematics.
The proposed computational framework is amenable to many modern \textit{ab initio} methods. Therefore, this work paves the way for extending this approach to increasingly complex nuclei.

The paper is organized as follows. In Sec.~\ref{sec:background} we review the formalism of response functions, integral transforms and the Chebyshev approach. Section~\ref{sec:binning} details the binning strategy using an estimation of the density of states. Section~\ref{sec:results} presents our results both for the density of state estimation as well as relevant electron-nucleus scattering responses at different kinematics. Finally, Sec.~\ref{sec:summary} offers our conclusions.

\section{Background}
 \label{sec:background}

Nuclear response functions characterize the transition probability of a nucleus in its ground state, $\ket{\Psi_0}$ with energy $E_0$, to a final state $\ket{\Psi_f}$ with energy $E_f$, under the action of an external probe. In the most general case, both energy $\omega$ and three-momentum $q = \abs{\mathbf{q}}$
 can be transferred from the probe to the nucleus. In the non-relativistic regime, however, the process is mediated by an operator that depends only on the momentum transfer $\hat{O}(q)$. The corresponding spin-averaged response function can be written as
\begin{equation}   \label{generalresponsefunctiontheory}
    R(q, \omega) = \frac{1}{2J_0 +1 } \SumInt \big| \bra{\Psi_f}  \hat{O}(q)  \ket{\Psi_0} \big|^2 \delta(E_f - E_0 - \omega),
\end{equation}
where the symbol $\SumInt$ denotes a sum over discrete states and an integral over continuum states. The factor $1/(2J_0 +1)$ averages over the $2J_0 + 1$ spin projections of the initial-state nucleus in its ground state of spin $J_0$, which we omit in what follows for simplicity.

In finite-dimensional model spaces, such as those used in numerical many-body calculations, the Hamiltonian spectrum determined by $E_f$ (and the corresponding $\ket{\Psi_f}$) is discrete by construction. Consequently, the response function of Eq.~\eqref{generalresponsefunctiontheory} will be discrete, even in a regime where it should physically be in continuum, e.g., after particle-disintegration threshold. 

To avoid computing a discretized response directly, one may employ integral transforms~\cite{efros2007,Carlson_1992,Lovato_2016} to 
recast the continuum problem to a bound-state like one. The integral transform of the response function $R(q, \omega)$ is defined as
\begin{align}
    \label{eq:integraltransform}
    \Phi(\sigma,q) & = \int d\omega\, K(\sigma, \omega) R(q, \omega) \\ \nonumber
    &  = \bra{\Psi_0} \hat{O}^\dagger({q}) K(\sigma, \hat{H} - E_0) \hat{O}({q}) \ket{\Psi_0},
\end{align}
where $K(\sigma, \omega)$ is a kernel function. Because the integral transform in Eq.~\eqref{eq:integraltransform} is written as the expectation value of an operator taken on the ground state $\ket{\Psi_0}$, it can be calculated with bound-state techniques. 

The main advantage of integral transforms is that they avoid explicit construction of the final states $\ket{\Psi_f}$, which in principle should be continuous, not discrete. As such, they are widely used in nuclear physics~\cite{Carlson_1992,efros2007,bacca2002, bacca2013, Lovato_2016,Sobczyk:2021dwm,Sobczyk:2023sxh}. 
Commonly used ones are, e.g., the Laplace transform~\cite{Carlson_1992}, the Lorentz integral transform~\cite{efros2007}, and the Gaussian integral transform~\cite{sobczykroggero}. 
However, to retrieve $R(q, \omega)$ from Eq.~\eqref{eq:integraltransform}, one has to invert the transform itself. This is a notoriously difficult problem. Reconstructing a response function from an integral transform (like Laplace or Lorentz transform) is ill-posed because small errors in the transform can cause large errors in the inverted response function. While this difficulty is largely under control~\cite{efros2007,Parnes:2025seu}, the transform inversion remains a source of numerical uncertainties that are not easily quantifiable.

In order to circumvent the inversion step, we follow the strategy laid out in Ref.~\cite{sobczykroggero} and applied in Refs.~\cite{Sobczyk:2022ezo,Sobczyk:2023mey, Sobczyk:2024hdl}, to reconstruct response functions in terms of a histogram using Chebyshev expansions. To this end, we first calculate the Gaussian integral transform (GIT) using the Gaussian kernel with a small width. 
This allows to resolve the discrete structure of the system with narrow peaks. Next, we bin these peaks with histograms.

In essence, we are interested in estimating a histogram of the response
function 
\begin{equation}
    h(\eta, \Delta) = \int d\omega R(q, \omega) f(\omega; \eta, \Delta),
\end{equation}
where $f(\omega; \eta, \Delta)$ is a function in $\omega$ with parameters $\eta, \Delta$. Specifically, $f$ is a window function of width $2\Delta$ centered in $\eta$ given as
\begin{equation}
    \label{eq:windowfunctiondefinition}
    f(\omega; \eta, \Delta) = 
    \begin{cases}
        1 & \text{if } |\eta - \omega| \leq \Delta, \\
        0 & \text{otherwise}.
    \end{cases}
\end{equation}
The histogram $h(\eta, \Delta)$ in principle averages the response function in a $2\Delta$ wide bin centered at $\eta$. We can approximate $h(\eta, \Delta)$ by $\tilde{h}(\eta,\Delta)$ using the integral transform (see Eq.~\eqref{eq:integraltransform})
\begin{equation}
    \tilde{h}(\eta,\Delta) =  \int d\sigma \,\Phi(\sigma, q) f(\sigma; \eta, \Delta).
    \label{eq:IT_hist}
\end{equation}
We calculate the integral transform $\Phi$ by expanding it into a set of orthogonal polynomials 
\begin{equation}
    \label{eq:chebyshevexpansion}
    \Phi(\sigma, q) = \int d\omega\, K(\omega, \sigma) R(q, \omega) = \sum_{k=0}^{\infty} c_k(\sigma)\, m_k,
\end{equation}
where the moments $m_k$ are defined as
\begin{align}
    \label{eq:moments}
    m_k &= \int d\omega\, T_k(\omega) R(q, \omega) \\ \nonumber
    &= \bra{\Psi_0} \hat{O}^\dagger(q)\, T_k(\hat{H}')\, \hat{O}(q) \ket{\Psi_0}.
\end{align}
Here, $T_k(\omega)$ are Chebyshev polynomials of the first kind, and $\hat{H}'$ is a rescaled Hamiltonian such that its spectrum lies within $[-1, 1]$, on which Chebyshev polynomials are defined. In practice, this expansion will be terminated at a finite number of moments $N_{\textrm{mom}}$.
We note that using this expansion, the kernel dependence in Eq.~\eqref{eq:IT_hist} is encoded in the coefficients $c_k$ while the moments $m_k$ are calculated independently.

By calculating the integral transform in terms of Chebyshev polynomials, the uncertainty bound of this procedure has an analytical form. First, we define a general property of the kernel to be \emph{$\Sigma$-accurate with $\Lambda$-resolution} if
\begin{equation}
    \inf_{\omega_0 \in (E_{\min}, E_{\max})} \int_{\omega_0 - \Lambda}^{\omega_0 + \Lambda} d\sigma\, K(\sigma, \omega_0) \geq 1 - \Sigma,
\end{equation}
which for a Gaussian kernel
\begin{equation}
    \label{eq:gaussian}
    K(\omega, \sigma) = \frac{1}{\sqrt{2\pi} \lambda} \exp\left[-\frac{(\sigma - \omega)^2}{2\lambda^2}\right],
\end{equation}
gives
\begin{equation}
    \Sigma \leq \exp\left(-\frac{\Lambda^2}{2\lambda^2}\right).
\end{equation}

The first source of uncertainty in reconstructing $h(\eta, \Delta)$ via $\tilde{h}(\eta,\Delta)$ is due to the finite kernel resolution
\begin{equation}
    \tilde{h}(\eta, \Delta - \Lambda) - m_0 \Sigma \leq h(\eta, \Delta) \leq \tilde{h}(\eta, \Delta + \Lambda) + m_0 \Sigma,
\end{equation}
where $m_0 = \int d\omega\, R(q, \omega)$ is the first Chebyshev moment or the total strength of the response, a ground-state property. Including the truncation error from terminating the Chebyshev expansion at $ N_{\textrm{mom}}$, the total reconstruction error is bounded by
\begin{align}
    \label{eq:totalbound}
    &\tilde{h}_{N_{\textrm{mom}}}(\eta, \Delta - \Lambda) - m_0 \Sigma - 2 m_0 \beta_{N_{\textrm{mom}}}(\Delta - \Lambda) 
    \leq h(\eta, \Delta), \\ \nonumber
    & h(\eta, \Delta) \leq \tilde{h}_{N_{\textrm{mom}}}(\eta, \Delta + \Lambda) + m_0 \Sigma + 2 m_0 \beta_{N_{\textrm{mom}}}(\Delta + \Lambda),
\end{align}
where $\beta_{N_{\textrm{mom}}}$ is a kernel-dependent bound related to truncation, and can be calculated analytically for the Gaussian case (see Eq.~(B10) in Ref.~\cite{sobczykroggero}). We would like to point out that this is \textit{purely} the reconstruction error, which does not include any model space, interaction or many-body uncertainty.

The error bounds in Eq.~(\ref{eq:totalbound}) quantify the reconstruction uncertainty of the (nuclear) response functions. They depend on several parameters. Some, like $\Lambda$, which relates to the kernel resolution $\lambda$, have a direct and intuitive impact on the uncertainty and their impact can be reduced with increased computational effort straightforwardly. Others, such as the sets of midpoints \(\{\eta_i\}\) and widths \(\{\Delta_i\}\), are free parameters that define the binning scheme used in the reconstruction. While these do not affect computational scaling, their choice significantly influences the reconstruction errors. Not all binning schemes are equally effective; thus, minimizing reconstruction error and maximizing predictivity requires a carefully designed binning strategy.

\section{Binning strategy}
\label{sec:binning}
To reconstruct a continuum response with controlled uncertainty using a Chebyshev expansion, our only freedom lies in choosing the histogram bins (parameters $\eta$ and $\Delta$ in Eq.~\eqref{eq:windowfunctiondefinition}). In a bound‐state framework the continuum appears as a finite set of discrete Hamiltonian eigenvalues, located along the energy‐transfer axis $\omega$, to which do not have a direct access. The GIT of the distribution of these eigenvalues guides our binning.

We draw inspiration from experimental histograms: discrete eigenvalues play the role of observed ``events'' and the number of eigenvalues per bin determines how well we discretize the continuum, analogous to a statistical uncertainty in experiment. If a bin is too narrow (contains no eigenvalues), it yields (almost) zero response and artificially redistributes spectral weight into adjacent bins as the total strength, the moment $m_0$, is conserved. If a bin is too wide, fine features (e.g.\ peaks) are smeared out. An effective binning avoids both extremes by ensuring each bin contains a similar, reasonable number of eigenvalues.

In our case, we want to get access to a regularized density of states (DOS) for our operator/response of interest, in which all accessible states equally contribute. In practice, the exact eigenvalue density is inaccessible, so we estimate a regularized DOS for $\hat{O}(q)$ via a stochastic calculation of the integral transform $\Phi(\sigma,q)$.

Our stochastic approach, described below, is in spirit similar to other DOS or random trace estimations~\cite{RePEc:wsi:ijmpcx:v:05:y:1994:i:04:n:s0129183194000842,RevModPhys.78.275,Cortinovis2022}.
Namely, the DOS of interest is constructed as an integral transform involving a randomized ground state $\ket{\Psi_0 }$ and operator $\hat{O}(q)$, where we only constrain the relevant quantum numbers of the product $\hat{O}(q) \ket{\Psi_0 }$. We then use this estimated DOS to guide the binning of the deterministic (non-randomized) response function. This procedure minimizes discretization errors and maximizes resolution without imposing any assumptions, relying solely on the structure of the accessible state space. In terms of binning, we partition the total $\omega$‐range into intervals of equal integrated DOS, ensuring approximately uniform ``event'' counts per bin while adapting to the underlying spectrum.

In order to reconstruct the DOS via the Chebyshev moments, let us start with the ``pivot'' state
\begin{equation}
\ket{\alpha} \equiv \hat O\,\ket{\Psi_0}\,,
\end{equation}
where we omit the explicit $q$–dependence of $\hat O$ for a simpler notation. 

Expanding the ground state in some orthonormal computational basis $\{\ket{i}\}$ as
\begin{equation}
\ket{\Psi_0} = \sum_i c_i\,\ket{i}, 
\quad
O_{ji} \equiv \bra{j}\hat O\ket{i},
\end{equation}
the pivot becomes
\begin{equation}
\ket{\alpha}
= \hat O\ket{\Psi_0}
= \sum_j \Bigl(\sum_i O_{ji}\,c_i\Bigr)\,\ket{j}
\;\equiv\;\sum_j \alpha_j\,\ket{j}.
\end{equation}
In the Hamiltonian eigenbasis $\{\ket{\Psi_n}\}$ one would instead write
\begin{equation}
\ket{\alpha} = \sum_n d_n\,\ket{\Psi_n},
\label{eq:alpha}    
\end{equation}
but obtaining all $\{\ket{\Psi_n}\}$ is in general computationally very hard. Therefore, we now turn to a stochastic sampling approach.
In order to estimate the DOS, our goal is to ensure that each eigenstate contributes equally to the Chebyshev moments in terms of their statistical expectation $\mathbb E$, so that 
$\mathbb E[\,|d_n|^2] \neq 0 $ and is constant for all $n$. Since the two bases are related by a unitary matrix $U$, \(d = U\,\alpha\), 
it suffices to replace the deterministic coefficients 
\(\alpha_j = \sum_i O_{ji}c_i\) 
by independent and identically distributed random variables $\xi_j$ drawn from a zero-mean, unit-covariance distribution, i.e.,
\begin{equation}
    \label{eq:distributionrequirements}
    \mathbb{E}[\xi_j] = 0, \quad \mathbb{E}[\xi_i \xi_j] = 0\, \, \forall i \neq j, \quad 
    \mathbb{E}[\xi_i^* \xi_j] = \delta_{ij}.
\end{equation}
The unitarity of $U$ guarantees that
\begin{align}
\mathbb{E}\bigl[|d_n|^2\bigr]
    &= \mathbb{E}\Bigl[\Bigl|\sum_i U_{n i}\,\xi_i\Bigr|^2\Bigr] \nonumber \\
    &= \sum_{i,j}U_{n i}\,U_{n j}^*
       \,\mathbb{E}\bigl[\xi_i\,\xi_j^*\bigr] \nonumber \\
    &= \sum_i \bigl|U_{n i}\bigr|^2 = 1,
\end{align}
for all $n$, i.e.\ equal average weight per eigenstate. This is exactly what we want in order to estimate the DOS relevant for our response function. We emphasize here that the randomized pivot $\ket{\xi}$ should carry the appropriate quantum numbers of the deterministic $\ket{\alpha}$ state in Eq.~\eqref{eq:alpha}. 

The convergence of our randomized approach is determined by how uniformly we sample the $|d_n|^2$ by drawing the $\xi_j $. This uniformness is determined by the fluctuation of $|d_n|^2$, $(\delta |d_n|^2)^2= \mathbb E[(|d_n|^2)^2] -(\mathbb E[|d_n|^2])^2 $ which is dominated by the fourth moment $\mathbb E[\xi_j^4]$. Since the $\xi_j $ are standardized (zero mean and unit variance), this fourth moment corresponds to the kurtosis, which characterizes the tail behavior and concentration of the distribution~\cite{Vershynin_2018}.
Several distributions satisfy Eq.~(\ref{eq:distributionrequirements}), including uniform draws on $[-1,1]$, standard Gaussian vectors, and the Rademacher distribution. All of these are subgaussian, such that their higher moments, e.g., the kurtosis are favorably bounded~\cite{Vershynin_2018}. We have numerically verified that all three distributions work, but Rademacher sampling—drawing each $\xi_j=\pm1$ with equal probability—is the most efficient choice because
\begin{itemize}
  \item $|\xi_j|=1$ exactly, so it is less likely that the $d_n$ are accidentally tiny compared to the other two distributions;
  \item $\mathbb{E}[|\xi_j|^4]=1$ for Rademacher versus $3$ for Gaussian, yielding more uniform sampling;
  \item  We sample a sign, so only one bit per sample is needed, avoiding the floating‐point cost of Gaussian or uniform draws.
\end{itemize}

As previously mentioned, we propose to use this regularized DOS to inform the binning strategy. We define bins by partitioning the total $\omega$ support of the response into intervals of equal integrated strength. These intervals correspond to our defined bins. Since each eigenstate contributes equally on average, a sufficiently large number of draws $N_{\text{draw}}$ ensures that regions with similar integrated density contain approximately the same number of eigenvalues. Moreover, assuming that we know the size of the discretized space, i.e., the dimension of the Hamiltonian, we can assure that each bin contains a sufficient number of eigenvalues.

The procedure is outlined in Algorithm~\ref{ag:dos}.
\begin{figure}[H]
\begin{algorithm}[H]
\caption{Estimate DOS and determine bins}
\label{ag:dos}
\begin{algorithmic}[1]
  \State Draw \(N_{\rm draw}\) random pivots \(\{\ket{{\xi^{(l)}}}\}\), where \(l = 1, \ldots, N_{\rm draw}\), from a zero‐mean, unit‐covariance ensemble.
  \State Compute and average the first \(N_{\rm mom}\) Chebyshev moments: 
    \(\overline{m}_k = \tfrac1{N_{\rm draw}}\sum_l m_k^{(l)}\).
  \State Reconstruct \(\Phi(\sigma)\) from \(\{\overline{m}_k\}\).
  \State Determine bin properties \(\{\eta_i\}, \{\Delta_i\}\) based on \(\Phi(\sigma)\) such that reconstruction error and resolution are appropriate.
\end{algorithmic}
\end{algorithm}
\end{figure}
This approach allows us to assess how well the continuum is discretized within each bin. In general, we can trade resolution for smaller errors by merging subsequent bins. When bins are merged, the ratio of the number of eigenvalues to bin width remains similar, implying that the discretization quality in merged bins is comparable to that in unmerged bins. 
This supports a hybrid strategy: merging bins in regions where the reconstruction errors are large, and preserving finer resolution where important physical features reside.

\section{Results}
\label{sec:results}
Having proposed a binning strategy to reconstruct response functions from the Chebyshev expansion, we now apply it to a realistic physics case. For this purpose, we consider the nuclear two-body problem, i.e., the deuteron, as a simple test bench. For such a system, calculations are efficient and numerically well controlled.
In fact, full diagonalization is computationally straightforward for the deuteron, making it an ideal testing ground for validating our strategy based on the regularized DOS and its stochastic reconstruction.

We use a Hamiltonian composed of the kinetic energy and a potential derived in chiral effective field theory at next-to-next-to-next-to leading order with a cutoff of 500$\,$MeV, as developed by Entem and Machleidt~\cite{Entem:2003ft}.
We choose this interaction because it is widely used in many-body systems, often supplemented by a three-body force.

The computational framework is based on a truncated harmonic oscillator (HO) basis expansion. This basis is widely used in nuclear theory and underlies many-body methods such as the no-core shell model~\cite{Barrett:2013nh} and coupled-cluster theory~\cite{Hagen:2013nca}, in-medium similarity renormalization group~\cite{Heiko}, and the self-consistent Green's function method~\cite{SCGF}.
Calculations in the HO basis depend on two parameters: the oscillator frequency \( \hbar\Omega \) and \( N_{\text{max}}\), i.e., the maximum quantum number \( N= 2n + l \), where $n$ is the radial quantum number and $l$ is the relative angular momentum between the proton and the neutron. The dependence of our results on these parameters will be discussed in detail. 

In the following, we benchmark our Chebyshev polynomial-based approach on two response functions of interest: the \gls{e1} response $R_{\rm E1}(\omega)$ and the longitudinal ($L$) response $R_L(\omega)$. The \gls{e1} response is comparatively simpler than the longitudinal response, as discussed below. Therefore, we use the \gls{e1} response to investigate the model space dependence and the quality of our binning strategy outlined in Sec.~\ref{sec:binning}. The model space dependence and binning quality for the longitudinal response are analogous and are therefore omitted from the following discussion. The response functions presented below are shown as histograms with associated reconstruction bounds: a lower bound (``Reconstruction low'') and an upper bound (``Reconstruction high''), as defined by the error bounds in Eq.~(\ref{eq:totalbound}).

\subsection{Electric dipole response (E1)}
As a first benchmark, we investigate the \gls{e1} response of the deuteron. This response is of interest as it provides the leading electric multipole contribution to the transverse response \( R_{\rm T}(q,\omega) \) in electron-nucleus scattering.
The deuteron photodissociation cross section \( \sigma_\gamma(\omega) \) reads
\begin{equation}
    \label{eq:photodissociation}
    \sigma_\gamma (\omega) = \frac{4 \pi^2 \alpha}{\omega} R_{\rm T}(q=\omega),
\end{equation}
and in the long-wavelength limit, \( q \to 0 \), it relates to the electric dipole response \( R_{\rm E1}(\omega) \) as~\cite{review}
\begin{equation}
    \label{eq:photodissociationE1}
    \sigma_\gamma (\omega) = 4 \pi^2 \alpha \omega R_{\rm E1}(\omega).
\end{equation}
Equation~\eqref{eq:photodissociationE1}, known as the unretarded dipole approximation, is a robust approximation for photodissociation cross sections at energies well below the pion production threshold~\cite{review} and allows for direct comparison between our calculated \( R_{\rm E1}(\omega) \) and experimental data in this \( \omega \) range.

\begin{figure}
    \centering
   
    \includegraphics[width=\columnwidth]{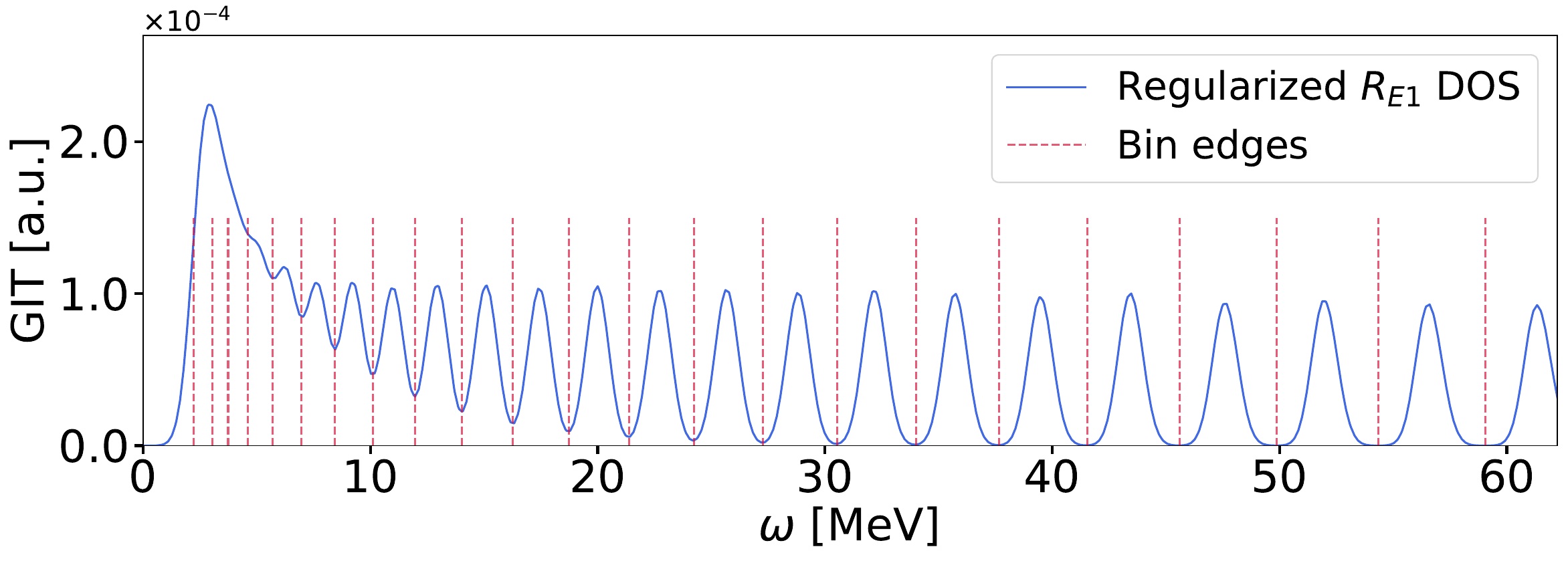}
    
    \caption{Binning determined by exact diagonalization of the full Hamiltonian and the regularized DOS for the \gls{e1} response ($N_{\textrm{max}} = 200, \hbar \Omega = 8 \ \text{MeV}$).}
    \label{fig:binningdos}
\end{figure}

We set the binning for this operator by calculating the regularized DOS, keeping the same quantum numbers of the randomized operator as for \gls{e1}. In Fig.~\ref{fig:binningdos} we show results of the randomized DOS estimation employing \( \lambda = 0.5\,\text{MeV} \) and \(N_{\text{draw}} = 2000 \) sampling from the Rademacher distribution. 
We observe that generally the eigenvalue density decreases as $\omega$ increases. On a finer scale, the regularized DOS periodically shows local minima corresponding to regions with small eigenvalue density in between regions of larger eigenvalue distribution. In the unbound regime, the Hamiltonian is dominated by kinetic energy, causing successive eigenvalues to differ primarily due to kinetic spacing. As a result, unbound eigenstates with different quantum numbers cluster around similar energies, with higher-lying states separated by discrete kinetic excitations. This also explains why the distance between local minima grows with energy $\omega$ and is proportional to the oscillator frequency $\hbar\Omega$. Local minima in the regularized DOS are ideal candidates for bin edges, as they minimize the reconstruction error. This can be understood from Eq.~(\ref{eq:totalbound}), where the uncertainty is mainly driven by a slightly smaller ($\Delta-\Lambda$) and slightly larger ($\Delta+\Lambda$) bin width. The difference of strength between the slightly smaller/larger histogram can be minimized when the bin edges are positioned in the region with the smallest contribution, i.e., in the local minima of the regularized DOS. As we are investigating the deuteron here, we can straightforwardly confirm this hypothesis by fully diagonalizing the system and determining the appropriate bin edges by grouping a constant number of eigenvalues—corresponding to the number of available channels—into each bin, and placing bin edges between the final eigenvalue of one group and the first of the next. These are shown by the red vertical lines in Fig.~\ref{fig:binningdos} and they coincide with the local minima of the regularized DOS. Therefore, we can reliably use this regularized DOS to determine the binning, with similar accuracy as the full diagonalization. We choose these minima as our bin edges.

The kernel resolution \( \lambda \) is crucial for determining the binning precision. For \( \lambda = 0.4\,\text{MeV} \) in Fig.~\ref{fig:binningdos}, the high density of eigenvalues at low \( \omega \) leads to substantial smearing by the kernel, resulting in significant overlap and the absence of clearly defined local minima in that region. 
We include this case to illustrate how larger kernel widths degrade resolution. As expected, reducing $\lambda$ decreases the overlap between smeared states and enables the visible bin separation in that region. In our subsequent results, we adopt significantly smaller values of $\lambda = O(50\,\mathrm{keV})$ to minimize resolution loss. 
If constraints on $\lambda$ exist for computational reasons, and in some $\omega$ regions local minima remain unresolved due to strong smearing, then the bin edges can alternatively be determined by enforcing equal area under the regularized DOS in each bin. This, in fact, would be the case of heavy many-body calculations. Importantly, we have verified that the area under the regularized DOS in between the red vertical lines in Fig.~\ref{fig:binningdos} is constant to the one percent level. This accuracy can be systematically improved by increasing $N_{\text{draw}}$ or decreasing $\lambda$, which in turn requires increasing $N_{\text{draw}}$ and $N_{\text{mom}}$.

For consistency, all results shown below use the same \( \lambda \) both for analyzing the DOS to set the binning and for reconstructing the response. If computational constraints impose a lower kernel resolution, then equal-area binning in the low-\( \omega \) region—as done in Fig.~\ref{fig:binningdos}—can result in large errors as the eigenvalues will not only contribute to a single bin. To alleviate this problem, one could merge adjacent bins, which effectively means that we double $\Delta$ keeping $\Lambda$ fixed in Eq.~\eqref{eq:totalbound}.

\begin{figure}
    \centering
   
    \includegraphics[width=\columnwidth]{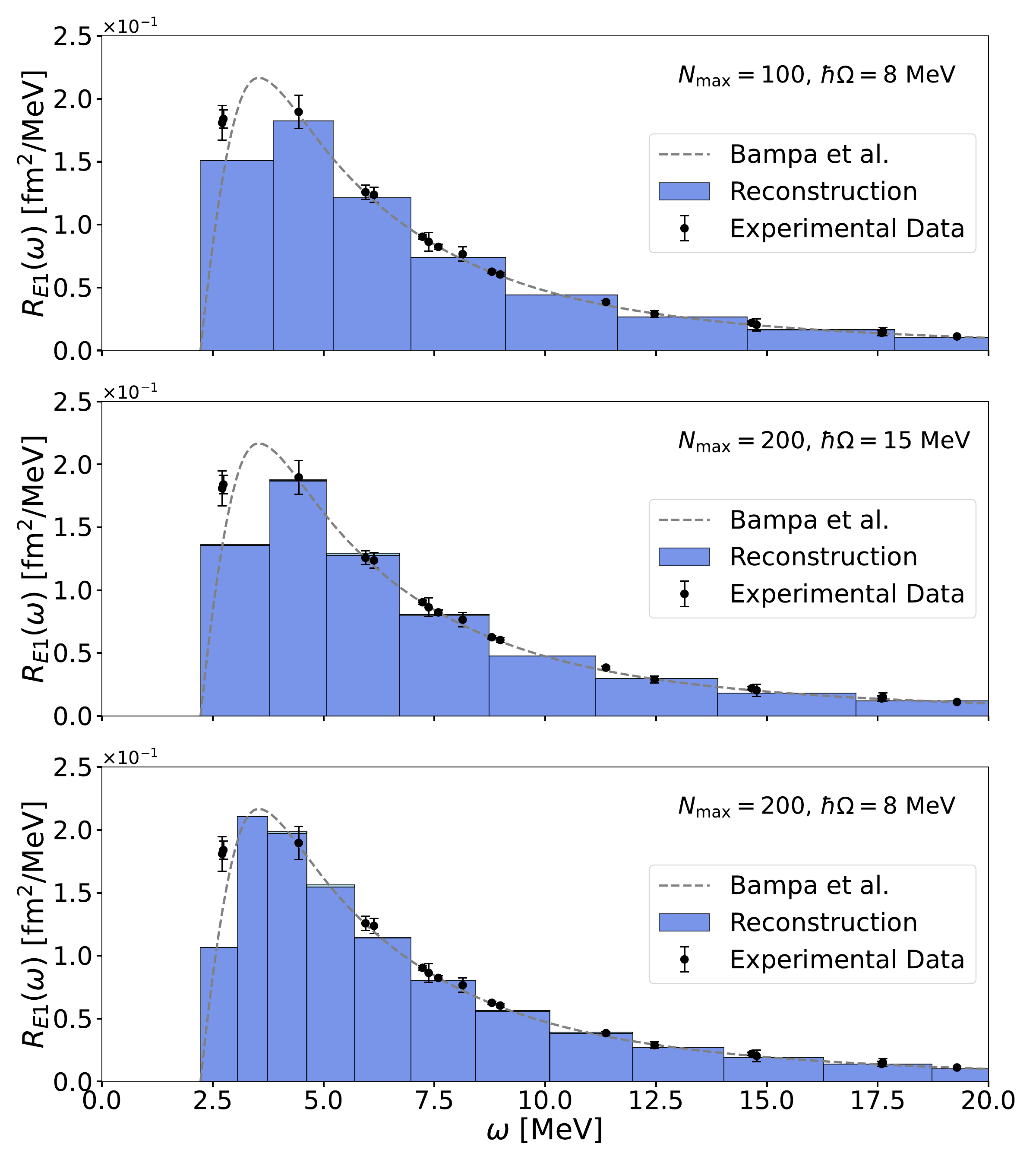}

    \caption{Comparison of the reconstructed response ($\lambda = 0.025\,$MeV, $N_{\textrm{mom}}=6000$), experimental data from Ref.~\cite{Arenhovel:1990yg} and the theoretical result of Ref.~\cite{Bampa:2011fq} for different model space parameters.}
    \label{fig:e1}
\end{figure}

Since we are focusing on a single multipole, we can systematically examine the model space dependence of our approach. 
In Fig.~\ref{fig:e1}, we present results for \( R_{E1}(\omega) \) for different values of \( N_{\text{max}} \) and \( \hbar\Omega \). The number of moments $N_{\textrm{mom}}$ and the kernel resolution $\lambda$ are constant here for all panels, and all results are well converged. We keep the same number of eigenvalues per bin in each panel.
We observe that decreasing \( \hbar\Omega \) at fixed \( N_{\text{max}} \) (second versus third panel) yields a finer resolution of the response. The same is true if \( N_{\text{max}} \) is increased for fixed \( \hbar\Omega \) (first versus third panel).
In the context of a HO basis, this dependence can be understood as follows: \( N_{\text{max}} \) sets the scale for the total number of eigenvalues, while the product \( N_{\text{max}} \cdot \hbar\Omega \) sets the upper energy scale of the Hamiltonian. Consequently, smaller \( \hbar\Omega \) at constant \( N_{\text{max}} \) leads to a denser distribution of eigenvalues as the energy range of the Hamiltonian shrinks. Increasing \( N_{\text{max}} \) increases the total number of eigenvalues, but also the energy range of the Hamiltonian and therefore of the eigenvalues. If both \( N_{\text{max}} \) and \( \hbar\Omega \) are increased by roughly the same factor of 2 (first and second panel), the resolution stays approximately constant. Although the energy range increases by a factor of 4—since it is proportional to \( N_{\text{max}} \cdot \hbar\Omega \)—the number of eigenvalues, which is proportional to \( N_{\text{max}} \), only doubles. Much of the additional range is dominated by kinetic energy, so the low-energy eigenvalue density changes little. The spacing there grows by a factor of 2 and therefore at low energies the density is similar. Hence, the eigenvalue density, which is relevant for the binning, is roughly constant in both panels.

To improve the binning resolution for relevant lepton-nucleus scattering observables in our approach we need to increase the number of low/medium energy eigenvalues. Therefore, decreasing \( \hbar\Omega \) is favored over increasing \( N_{\text{max}} \) to improve the binning resolution.
However, achieving ultraviolet convergence at lower \( \hbar\Omega \) typically requires a larger \( N_{\text{max}} \), which increases the computational cost~\cite{Furnstahl:2012qg}. Since \( N_{\text{max}} \) determines the dimensionality of the Hamiltonian matrix, all associated structures, such as wavefunctions, operator representations, and related quantities, scale accordingly. Maintaining comparable spectral resolution $\lambda$ generally requires increasing the number of Chebyshev moments if the Hamiltonian's total spectral range grows, since $N_{\text{mom}}$ depends on the ratio $\lambda / (\hbar \Omega N_{\text{max}})$ rather than purely on $\lambda$. 

When comparing to experimental data and to the theoretical result of Ref.~\cite{Bampa:2011fq}, performed with a realistic AV18 interaction, in Fig.~\ref{fig:e1}, we observe that in all three panels the binning reconstruction reproduces data very well and reconstruction uncertainties are negligibly small, i.e. the lower bound and upper bound of each bin basically coincide for all model spaces shown. As the eigenvalue density increases (i.e., as the discretization of the continuum improves), our binned curve approaches the smooth result of Ref.~\cite{Bampa:2011fq}. Both theoretical curves show a slight deviation from the first two experimental points, which is expected because magnetic-dipole contributions, which dominate near threshold, are neglected (see, e.g., Ref.~\cite{Arenhovel:1990yg}). When the results of Ref.~\cite{Bampa:2011fq} are averaged within the first bin, they are compatible with our binned reconstruction, confirming consistency in the threshold region.

\subsection{Longitudinal response}
Next, we focus on the longitudinal response $R_{\rm L}(q,\omega)$, which, together with $R_{\rm T}(q,\omega)$, describes quasi-elastic electron-nucleus scattering~\cite{review}. In spherical basis expansion methods, this quantity is usually calculated as an expansion of different multipoles. It is a more complex quantity than $\sigma_\gamma (\omega)$ in terms of $R_{E1}(\omega)$, because it involves contributions from various multipolarities along with an explicit $q$ dependence of the operator. 
In the following we settle on \( N_{\text{max}} =200 \) and \( \hbar\Omega = 8\,\)MeV as this generates sufficient resolution for all quantities we investigate.

We begin by analyzing the response at fixed $q$ and compare our results with those from the literature. In particular, we refer to two different sets: a low $q$ calculation at 20 MeV performed in \gls{pionless} in Ref.~\cite{Emmons:2020aov} and a high $q$ calculation (441$\,$MeV) performed with a realistic AV14-type interaction in Ref.~\cite{Efros:1993xy}. Both calculations are performed in the center-of-mass frame of the final state neutron-proton pair. We calculate multipoles 0 through 4 for $q=20\,$MeV and 0 through 6 for $q=441\,$MeV. At low $q$, fewer multipoles suffice, as the pionless calculation omits the expansion altogether, whereas the high‑$q$ result of Ref.~\cite{Efros:1993xy} likewise includes multipoles up to 6.

\begin{figure}
    \centering
    
    \includegraphics[width=\columnwidth]{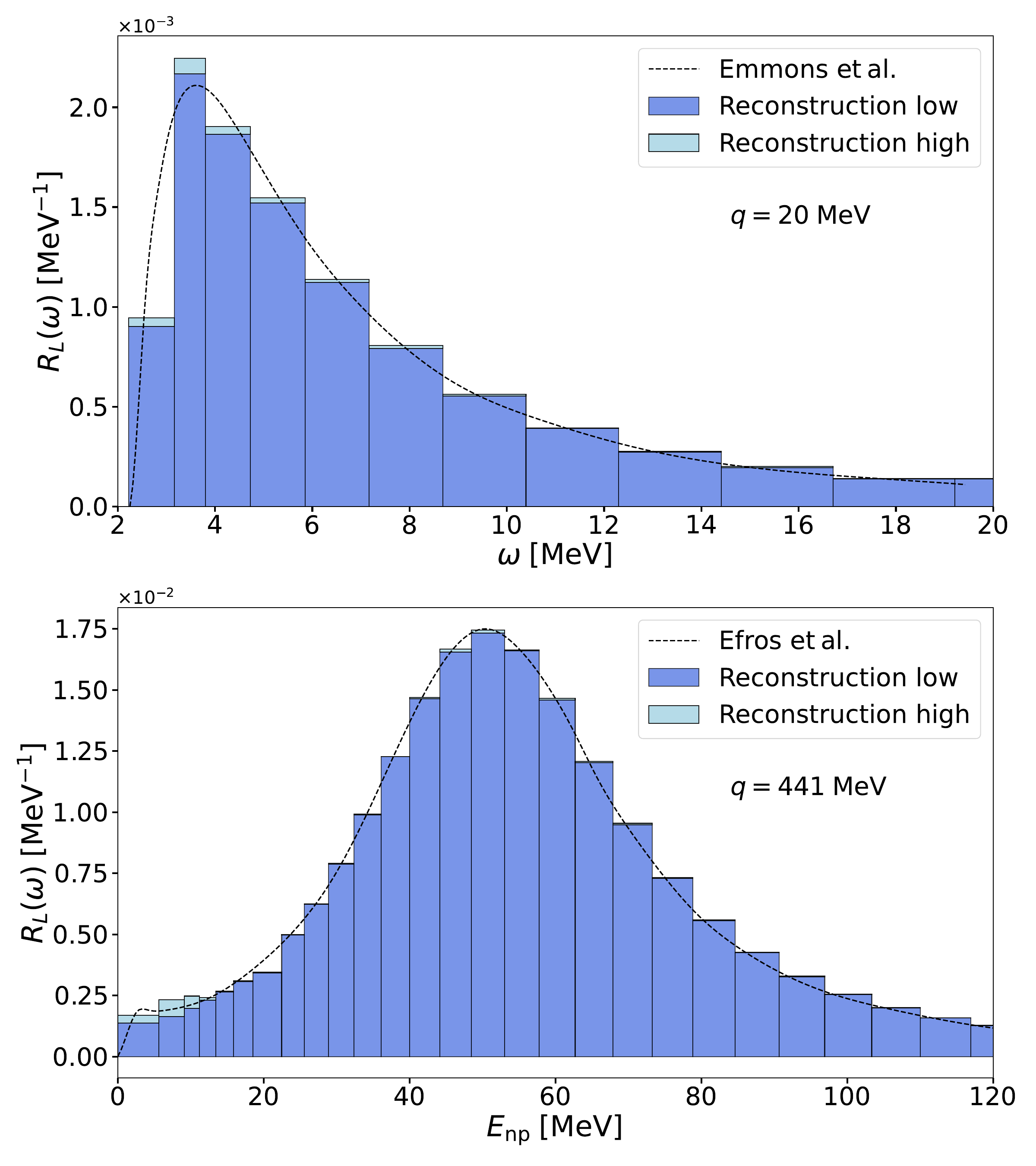}

    \caption{Calculation of the longitudinal response $R_{\rm L}(q,\omega)$ at different $q$ values and comparison with calculations from Refs.~\cite{Emmons:2020aov,Efros:1993xy} with multipoles 0 through 4 and 6, respectively ($N_{\textrm{max}} = 200$, $\hbar \Omega = 8 \,$MeV, $\lambda = 0.025/0.05\,$MeV, $N_{\textrm{mom}}=6000$).}
    \label{fig:longitudinalfixedq}
\end{figure}

As can be observed in Fig.~\ref{fig:longitudinalfixedq}, our results show excellent agreement with theoretical calculations across different $q$ regimes. In both cases reconstruction errors are typically largest in the low-energy regime just above threshold, where eigenvalues are densely packed. For $q = 441\,$MeV, we merge adjacent bins if the bin width is less than $2$ MeV; this reduces the error bars and clarifies the broad peak. In contrast, in the low‐$q$ regime, merging bins near threshold would obscure the sharp peak, so we retain the finer binning there. Likewise, because the main feature at high $q$ is broad, we adopt a coarser kernel resolution $\lambda=0.05\,$MeV, compared to $\lambda=0.025\,$MeV in the low $q$ case, to accelerate the computation. This choice does introduce somewhat larger errors at low $E_{\rm np}$, but these too could be mitigated—at the expense of resolution—by further bin merging.

Finally, we validate our approach by comparing to experimentally extracted $R_{\rm L}(\omega, q)$ values from Ref.~\cite{deuteronlongitudinalbates}. The Rosenbluth separation in Ref.~\cite{deuteronlongitudinalbates} has been performed for fixed values of electron energy $E$ and scattering angle $\theta$. As a result, momentum and energy in the laboratory frame $q^{\text{lab}}$ and $\omega^{\text{lab}}$ appear in fixed pairs as
\begin{equation}
    q^{\text{lab}}(\omega^{\text{lab}}) = \sqrt{E^2 + (E - \omega^{\text{lab}})^2 - 2E(E - \omega^{\text{lab}})\cos\theta}\,.
\end{equation}

When comparing our calculations with the $R_{\rm L}^{\text{lab}}(\omega^{\text{lab}}, q^{\text{lab}})$ measured in the laboratory frame, we need to make the following considerations. First, we do not calculate $R_{\rm L}^{\text{lab}}(\omega^{\text{lab}}, q^{\text{lab}})$ but rather the response function in the center-of-mass frame $R_{\rm L}^{\text{cm}}(E_{\textrm{np}}^{\text{cm}}, q^{\text{cm}})$, where $E_{\textrm{np}}^{\text{cm}}$ is the relative energy of the neutron proton pair in their center-of-mass frame after the breakup\footnote{To avoid confusion we now label all center-of-mass variables explicitly, i.e., what was $q$ before is now $q^{\text{cm}}$ and so on. }. Secondly, in our framework we calculate $R_{\rm L}^{\text{cm}}(E_{\textrm{np}}^{\text{cm}}, q^{\text{cm}})$ for a fixed $q^{\text{cm}}$ as a function of $E_{\textrm{np}}^{\text{cm}}$.

Since the experimental data are given in the laboratory frame while our calculations are performed in the center-of-mass frame, a Lorentz transformation is required to enable meaningful comparison.
Following Refs.~\cite{fabianarenhövel, christlmeiergrießmann} we relate $(\omega^{\text{lab}},q^{\text{lab}})$ to $(\omega^{\text{cm}},q^{\text{cm}})$ using a Lorentz boost along the momentum transfer $\mathbf{q}$. The relevant boost quantities $\beta, \gamma$ are given as \cite{christlmeiergrießmann}
\begin{equation}
    \beta = \frac{q^{\text{lab}}}{M_d + \omega^{\text{lab}}}, \gamma = \frac{1}{1-\beta^2},
\end{equation}
where $M_d$ is the deuteron mass. The experimentally measured kinematical quantities then transform as
\begin{equation}
    \omega^{\text{cm}} = \gamma \omega^{\text{lab}} - \beta \gamma q^{\text{lab}},\, q^{\text{cm}} = \beta \gamma M_d.
\end{equation}

The final state relative energy $E_{\text{np}}^{\text{cm}}$ accounts for the recoil of the deuteron and serves as the relevant energy variable in our cm-frame calculation~\cite{fabianarenhövel}
\begin{equation}
    E_{\textrm{np}}^{\text{cm}} = \omega^{\text{cm}} + \sqrt{(q^{\text{cm}})^2 + M_d^2} - 2 M.
\end{equation}
Following Refs.~\cite{Beck:1992yd,Efros:2005uk}, the responses in the two frames are related according to
\begin{equation}
    R_L^{\text{lab}}(\omega^{\text{lab}}, q^{\text{lab}}) = \frac{({q^{\text{lab}}})^2}{(q^{\text{cm}})^2} \frac{E_i^{\text{cm}}}{M_d} R_L^{\text{cm}}(E_{\textrm{np}}^{\text{cm}}, q^{\text{cm}}),
\end{equation}
where $E_i^{\text{cm}}=\sqrt{(q^{\text{cm}})^2 + M_d^2} $ is the initial state energy of the deuteron in the np final state center-of-mass frame.

While either frame could be used for the comparison, we choose the center-of-mass frame, since the eigenvalue binning is independent of $q^{\text{cm}}$, whereas it would vary with $q^{\text{lab}}$ after transformation. To obtain a meaningful comparison to experimental data, we define the error as the sum of the reconstruction error and the spread of results obtained between the lowest and highest $q^{\text{cm}}$ values within each bin.

\begin{figure}
    \centering
    \includegraphics[width=\columnwidth]{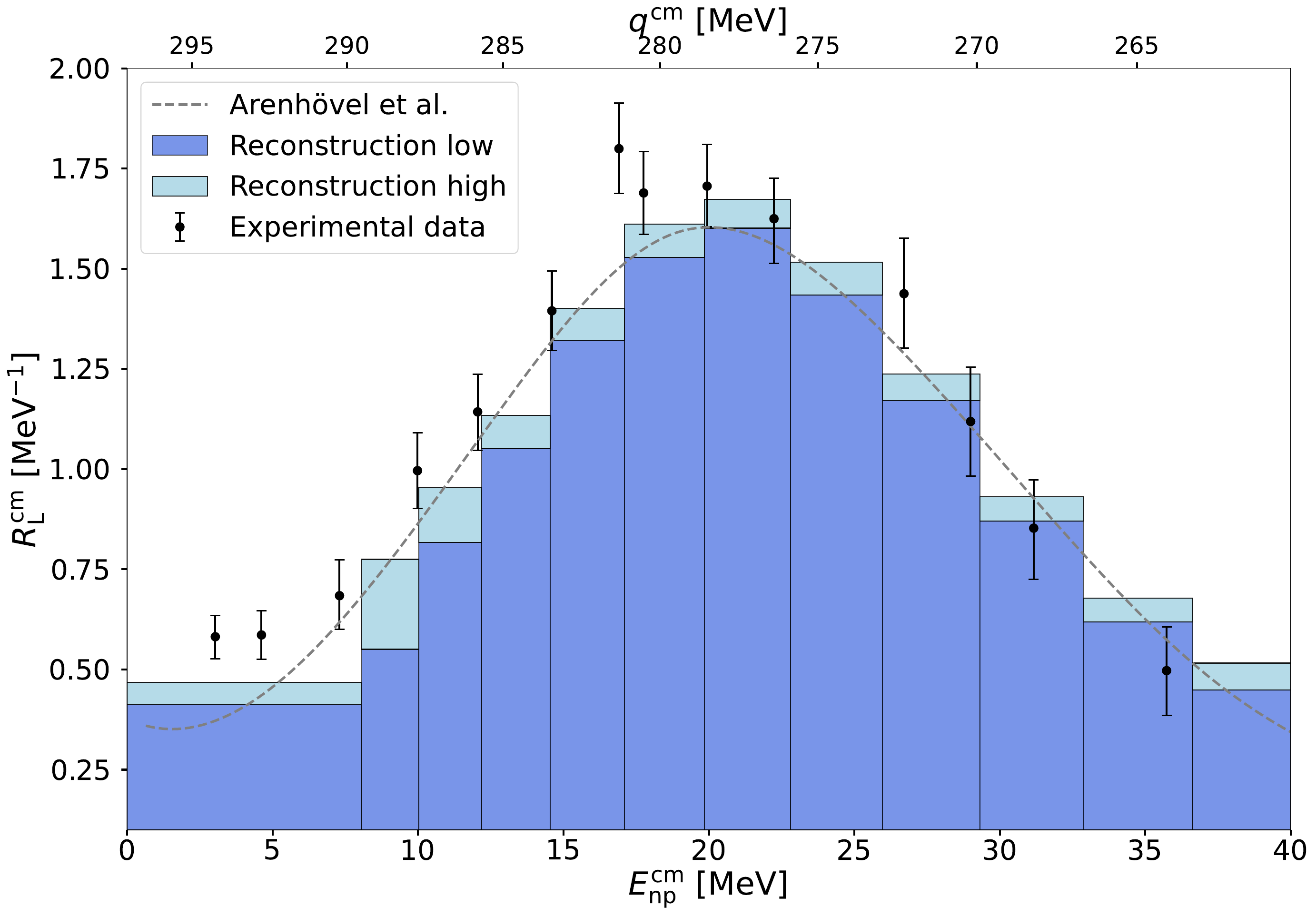}
    \caption{Comparison of the reconstructed longitudinal response ($N_{\textrm{max}}=200$, $\hbar \Omega=8\,$MeV) to experimental data and a calculation, both found in Ref.~\cite{deuteronlongitudinalbates}.}
    \label{fig:longitudinalvaryingq}
\end{figure}

In Fig.~\ref{fig:longitudinalvaryingq}, we show our results for $R_{\rm L}^{\text{cm}}(E_{\textrm{np}}^{\text{cm}}, q^{\text{cm}})$, the experimental data and a theoretical calculation by Arenh\"ovel and Leidemann, both from Ref.~\cite{deuteronlongitudinalbates}, considering multipoles from $0$ through $7$ and taking the dipole parametrization of the nucleon form factors~\cite{HOHLER1976505}. We observe very good agreement with both the data and the calculation found in Ref.~\cite{deuteronlongitudinalbates}, which is a further proof of the robustness of the proposed method. The relatively large error bars compared to Fig.~\ref{fig:longitudinalfixedq} arise from the additional uncertainty associated with the variation of $q^{\text{cm}}$ within each bin.

\section{Conclusions}
\label{sec:summary}
We have presented the first systematic implementation of a Chebyshev‐histogram integral‐transform approach to electromagnetic response functions in finite nuclei that circumvents the ill-posed inversion step.
Within a bound‑state framework, we treat the discretized eigenvalue spectrum as a proxy of the continuum, and devise a physically motivated binning strategy. We highlighted the importance of constructing a sensible binning strategy. We adaptively choose bins of equal eigenvalue count per unit width, placing edges at DOS minima to minimize reconstruction error.

To guide the binning, we introduced a stochastic method to estimate a regularized density of states using Chebyshev polynomials. By drawing random vectors and averaging over many samples, we obtain a smoothed, regularized eigenvalue distribution without having to resort to full diagonalization. This method can identify regions with low and high eigenvalue density, which are typical in bound-state methods. By integrating the curve, we can define equal-area bins with a similar number of eigenvalues. This allows for the reconstruction of the response using physically motivated bins that minimize the reconstruction error without requiring additional smoothness assumptions.

We benchmarked our approach using the electric dipole response function and the longitudinal response of the deuteron across various momentum transfers, finding excellent agreement with both theoretical calculations and experimental data.
The proposed approach is designed to be conceptually and computationally general, and is expected to apply across a wide range of nuclear interactions, many-body systems, and response functions.
Therefore, it paves the way for possible extensions to general lepton-nucleus cross sections in medium-mass nuclei, relevant to long-baseline neutrino experiments.

\begin{acknowledgments}
We are grateful to Weiguang Jiang and Alessandro Roggero for many insightful discussions and to Winfried Leidemann for valuable correspondence. ICR acknowledges financial support from the Mainz Physics Academy during part of the period in which this research was carried out. This work was supported in part by the Deutsche Forschungsgemeinschaft (DFG) through the Cluster of Excellence ``Precision Physics, Fundamental Interactions, and Structure of Matter'' (PRISMA${}^+$ EXC 2118/1) funded by the DFG within the German Excellence Strategy (Project ID 39083149) and by the DFG—Project No. 514321794 (CRC1660:
Hadrons and Nuclei as discovery tools).

\end{acknowledgments}

\bibliography{references}
\end{document}